\documentclass{svjour3}                     
\smartqed  
\usepackage{graphicx}
\begin{document}

\title{Neutron Star Properties in the Chiral Quark-Meson Coupling Model
}
\subtitle{
}


\author{
Tsuyoshi Miyatsu
\and
Koichi Saito
}


\institute{
Tsuyoshi Miyatsu
\at
Department of Physics,
Soongsil University, \\
Seoul 156-743,
Republic of Korea \\
\email{tmiyatsu@ssu.ac.kr}           
\and
Koichi Saito
\at
Department of Physics, Faculty of Science and Technology, \\
Tokyo University of Science (TUS), Noda 278-8510, Japan \\
\email{koichi.saito@rs.tus.ac.jp}
}

\date{Received: date / Accepted: date}

\maketitle

\begin{abstract}
We study the properties of neutron star using the chiral quark-meson coupling model, in 
which the quark-quark hyperfine interaction due to the exchanges of gluon and pion based on chiral symmetry is considered. 
We also examine the effects of hyperons and $\Delta$-isobars in a neutron star. 
Extending the SU(6) spin-flavor symmetry to more general SU(3) flavor symmetry in the vector-meson couplings to baryons, 
the maximum mass of neutron star can reach the recently observed, massive pulsar mass, 
$1.97 \pm 0.04 M_{\odot}$.
In this calculation, $\Lambda$ and $\Xi$ are generated in a neutron star, 
while $\Sigma$ and $\Delta$-isobars do not appear. 
\end{abstract}

\section{Introduction}
\label{intro}
Neutron stars are laboratories for dense matter physics, 
because they are composed of the densest form of hadrons and leptons. 
However, the detail of neutron star properties, 
for example, the mass, radius and the particle fractions in the core of neutron star, 
is not yet theoretically understood.  Thus, the 
observation of the mass and/or radius 
can provide strong constraints on the equation of state (EOS) of highly dense matter. The 
typical mass of a neutron star is known to be around $1.4 M_{\odot}$ (solar mass). 
However, a few massive neutron stars have recently been observed precisely. 
Shapiro delay measurements from radio timing observations of the binary millisecond pulsar (PSR J1614-2230) 
have indicated the mass of $1.97 \pm 0.04 M_{\odot}$~\cite{Demorest:2010bx}.  

However, it is difficult to explain the mass of PSR J1614-2230 by the EOS which has been calculated so far, 
because the inclusion of hyperons makes the EOS very soft and the maximum mass of a neutron star is thus reduced. 
Therefore, it is very urgent to reconcile this discrepancy.
 
\section{Chiral Quark-Meson Coupling Model}
\label{sec:CQMC}

In Ref.~\cite{Nagai:2008ai}, we have extended the quark-meson coupling (QMC) model~\cite{Saito:2005rv} 
to include the quark-quark hyperfine interaction 
due to the exchanges of gluon and pion based on chiral symmetry. 
We call this improved version the chiral QMC (CQMC) model.
Then, we have applied it to a neutron star within relativistic Hartree-Fock (RHF) approximation, and 
shown that the mass of PSR J1614-2230 can be explained~\cite{Miyatsu:2011bc,Katayama:2012ge}. 

In the present calculation, instead of RHF, we study the neutron-star properties in relativistic Hartree approximation. 
The Lagrangian density is chosen to be 
\begin{eqnarray}
	\mathcal{L} &=& \sum_{B}\bar{\psi}_{B}\left[ 
	i\gamma_{\mu}\partial^{\mu}
	- M_{B}^{\ast}(\sigma,\sigma^{\ast})
	- g_{\omega B}\gamma_{\mu}\omega^{\mu}
	- g_{\phi B}\gamma_{\mu}\phi^{\mu}
	- g_{\rho B}\gamma_{\mu}\vec{\rho}^{\mu}\cdot\vec{I}_{B} \right]\psi_{B}
	\nonumber \\
	&+& \frac{1}{2} \left( \partial_{\mu}\sigma\partial^{\mu}\sigma-m_{\sigma}^{2}\sigma^{2} \right)
	+ \frac{1}{2} \left( \partial_{\mu}\sigma^{\ast}\partial^{\mu}\sigma^{\ast}
	- m_{\sigma^{\ast}}^{2}\sigma^{\ast2} \right)
	\nonumber \\
	&+& \frac{1}{2}m_{\omega}^{2}\omega_{\mu}\omega^{\mu}- \frac{1}{4}W_{\mu\nu}W^{\mu\nu}
	+ \frac{1}{2}m_{\phi}^{2}\phi_{\mu}\phi^{\mu}- \frac{1}{4}P_{\mu\nu}P^{\mu\nu}
	\nonumber \\
	&+& \frac{1}{2}m_{\rho}^{2}\vec{\rho}_{\mu}\cdot\vec{\rho}^{\mu}
	- \frac{1}{4}\vec{R}_{\mu\nu}\cdot\vec{R}^{\mu\nu}
	+ \sum_{\ell}\bar{\psi}_{\ell}\left[i\gamma_{\mu}\partial^{\mu}-m_{\ell}\right]\psi_{\ell},
\end{eqnarray}
where
\begin{equation}
	W_{\mu\nu} = \partial_{\mu}\omega_{\nu} - \partial_{\nu}\omega_{\mu}, 
	\hspace{0.5cm}
	P_{\mu\nu} = \partial_{\mu}\phi_{\nu} - \partial_{\nu}\phi_{\mu}, 
	\hspace{0.5cm}
	\vec{R}_{\mu\nu} = \partial_{\mu}\vec{\rho}_{\nu} - \partial_{\nu}\vec{\rho}_{\mu},
	\label{eq:covariant-derivative}
\end{equation}
with $\psi_{B (\ell)}$ the baryon (lepton) field, $M_{B} (m_{\ell})$ the free baryon (lepton) mass 
and $\vec{I}_B$ the isospin matrix for baryon $B$. 
The sum $B$ runs over the octet baryons and $\Delta$-isobars, 
while the sum $\ell$ is for the leptons, $e^{-}$ and $\mu^{-}$. 
The baryon-baryon interaction is given through the exchanges of not only the 
isoscalar ($\sigma$ and $\omega$) and isovector ($\vec{\rho}$) mesons 
but also the isoscalar, strange ($\sigma^{\ast}$ and $\phi$) mesons. 
The effective baryon mass, $M_{B}^{\ast}$, 
is calculated by using the simple QMC or CQMC model, 
and it can be parameterized as a function of $\sigma$ and $\sigma^{\ast}$:
\begin{equation}
	M_{B}^{\ast}(\sigma,\sigma^{\ast})
	= M_{B} - g_{\sigma B}(\sigma)\sigma - g_{\sigma^{\ast}B}(\sigma^{\ast})\sigma^{\ast},
\end{equation}
with 
\begin{equation}
	g_{\sigma B}(\sigma) = g_{\sigma B}b_{B}\left[1-\frac{a_{B}}{2}\left(g_{\sigma N}\sigma\right)\right],
	\hspace{0.4cm}
	g_{\sigma^{\ast}B}(\sigma^{\ast}) 
	= g_{\sigma^{\ast}B}b_{B}^{\prime}
	\left[1-\frac{a_{B}^{\prime}}{2}\left(g_{\sigma^{\ast}\Lambda}\sigma^{\ast}\right)\right],
	\label{eq:scalar-coupling-constants}
\end{equation}
where $g_{\sigma N}$ and $g_{\sigma^{\ast}\Lambda}$ are respectively 
the $\sigma$-$N$ and $\sigma^{\ast}$-$\Lambda$ coupling constants at zero density. 
Here, we introduce four parameters, $a_{B}$, $b_{B}$, $a_{B}^{\prime}$ and $b_{B}^{\prime}$, 
to describe the mass, and these values are listed in Table~\ref{tab:1}. 
\begin{table}
\caption{Values of $a_{B}$, $b_{B}$, $a_{B}^{\prime}$ and $b_{B}^{\prime}$ 
for various baryons in the QMC or CQMC model. 
In this study, we assume that the strange mesons do not couple to the nucleon and $\Delta$-isobars.
}
\label{tab:1}
\begin{tabular}{lcccccccc}
\hline\noalign{\smallskip}
\         & \multicolumn{4}{c}{QMC} & \multicolumn{4}{c}{CQMC} \\
$B$       & $a_{B}$~(fm) & $b_{B}$ & $a_{B}^{\prime}$~(fm) & $b_{B}^{\prime}$ 
          & $a_{B}$~(fm) & $b_{B}$ & $a_{B}^{\prime}$~(fm) & $b_{B}^{\prime}$ \\
\noalign{\smallskip}\hline\noalign{\smallskip}
$N$       & 0.179 & 1.00 & ---   & ---  & 0.118 & 1.04 & ---   & ---  \\
$\Lambda$ & 0.172 & 1.00 & 0.220 & 1.00 & 0.122 & 1.09 & 0.290 & 1.00 \\
$\Sigma$  & 0.177 & 1.00 & 0.223 & 1.00 & 0.184 & 1.02 & 0.277 & 1.15 \\
$\Xi$     & 0.166 & 1.00 & 0.215 & 1.00 & 0.181 & 1.15 & 0.292 & 1.04 \\
$\Delta$  & 0.200 & 1.00 & ---   & ---  & 0.197 & 0.89 & ---   & ---  \\
\noalign{\smallskip}\hline
\end{tabular}
\end{table}
If we set $a_{B}=0$ and $b_{B}=1$, $g_{\sigma B}(\sigma)$ is identical to the $\sigma$-$B$ coupling constant 
in Quantum Hadrodynamics. 
This is also true of the coupling of $g_{\sigma^{\ast}B}(\sigma^{\ast})$. 

\section{SU(3) Symmetry and Neutron Stars}
\label{sec:SU(3)-symmetry}

To study the role of hyperons (Y) on the properties of neutron star, 
it is very important to extend SU(6) symmetry based on the quark model 
to the more general SU(3) flavor symmetry. 
In general, the interaction Lagrangian density in SU(3) is given by~\cite{Rijken:2010zzb} 
\begin{equation}
	\mathcal{L}_{int} = -g_8 \sqrt{2} \left[ \alpha {\rm Tr}([{\bar B}, M_8] B) 
	+ (1-\alpha ) {\rm Tr}(\{{\bar B}, M_8\} B) \right] 
	- g_1 \frac{1}{\sqrt{3}} {\rm Tr}({\bar B}B){\rm Tr}(M_1) , 
	\label{eq:SU(3)-Lagrangian}
\end{equation}
where $g_1$ and $g_8$ are respectively the coupling constants for the meson singlet and octet states, 
and $\alpha$ ($0 \leq \alpha \leq 1$) denotes the $F/(F+D)$ ratio. 
For details, see the reference~\cite{Rijken:2010zzb}.  

\begin{figure}[b]
  \includegraphics[width=132.5pt,keepaspectratio,clip,angle=270]{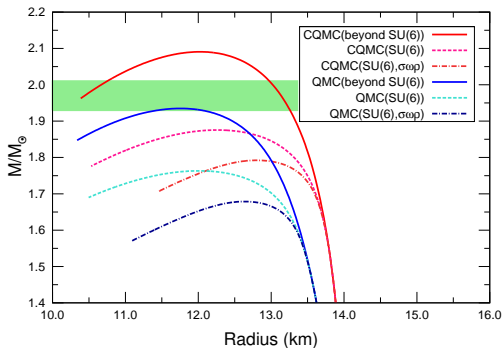}%
\caption{Mass-radius relation in the QMC or CQMC model.  The dot-dashed curves are for the SU(6) case without the strange mesons. 
The shaded area shows the mass of PSR J1614-2230, $1.97 \pm 0.04 M_{\odot}$~\cite{Demorest:2010bx}.
}
\label{fig:TOV}       
\end{figure}
The latest extended soft-core (ESC) model by the Nijmegen group suggests that 
the ratio $z(=g_{8}/g_{1})$ is 0.1949 for vector mesons, 
while $z=1/\sqrt{6}=0.4082$ in SU(6) symmetry. 
In the ESC model, the mixing angle is chosen to be $37.50^\circ$, which is close to the ideal mixing ($35.26^\circ$). 
Therefore, in the present study, we use the $z$ value in the ESC model 
and assume the ideal mixing to determine the vector-meson couplings to hyperons. 
For the $\sigma$-Y coupling constants, as in \cite{Miyatsu:2011bc,Katayama:2012ge}, 
we fix them so as to fit the hyperon potential suggested by the experiments. 

In Fig.~\ref{fig:TOV}, 
we present the mass-radius relation for a neutron star, which is calculated using 
the Tolman-Oppenheimer-Volkoff (TOV) equation with the EOS in the QMC or CQMC model. 
The inclusion of the strange mesons ($\sigma^{\ast}$ and $\phi$) makes the EOS stiff, and 
the maximum mass of neutron star is thus pushed upwards in both models. 
In fact, this enhancement is mainly caused by $\phi$ meson, and the effect of $\sigma^{\ast}$ meson is not large. 
Furthermore, extending SU(6) to SU(3) symmetry in the vector-meson couplings 
(this case is denoted by ``beyond SU(6)" in the figure), 
the maximum mass 
can reach the massive pulsar mass, $1.97 \pm 0.04 M_{\odot}$~\cite{Demorest:2010bx}. 

\begin{figure}
\includegraphics[width=98.2pt,keepaspectratio,clip,angle=270]{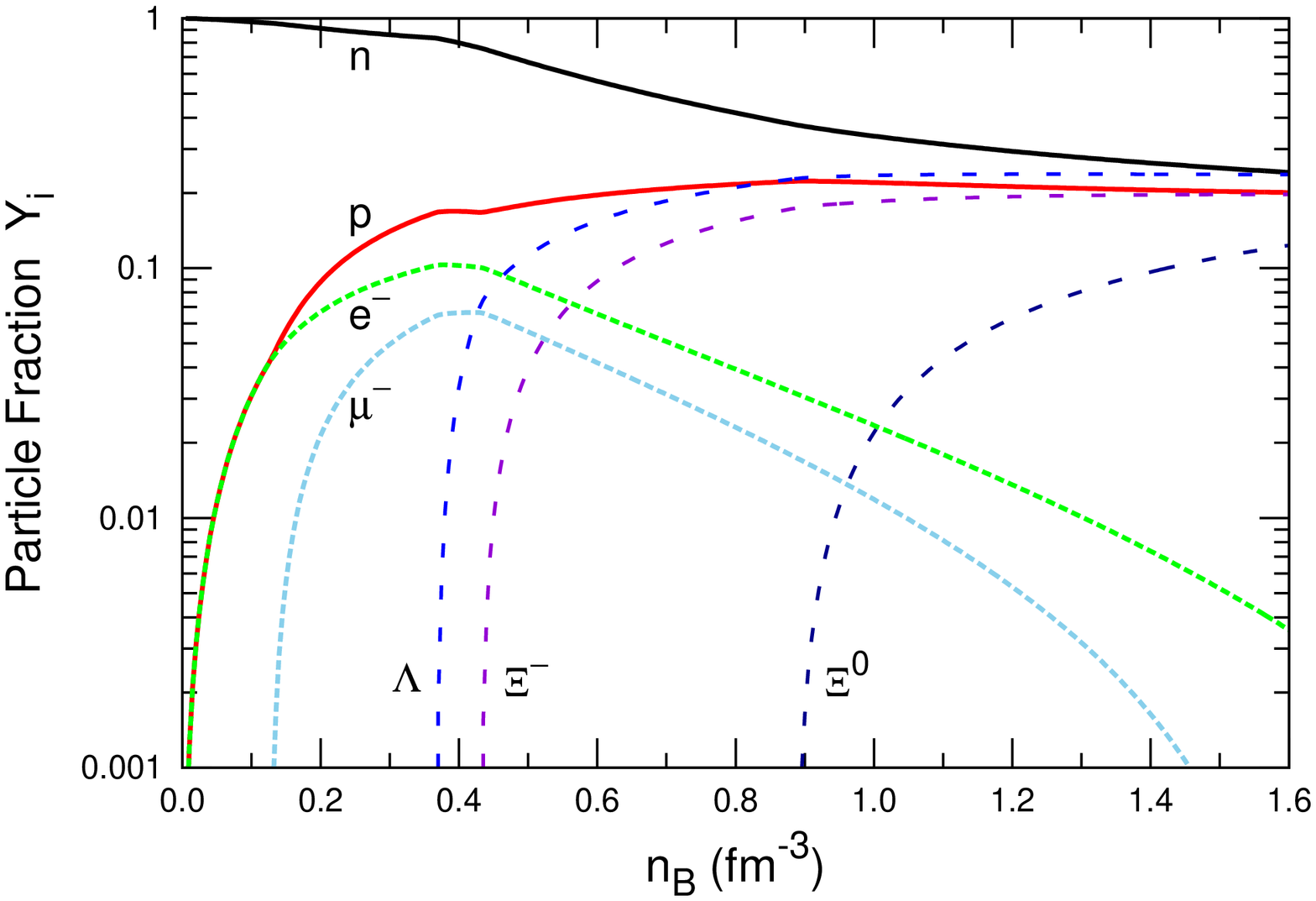}%
\hspace{0.6cm}
\includegraphics[width=98.2pt,keepaspectratio,clip,angle=270]{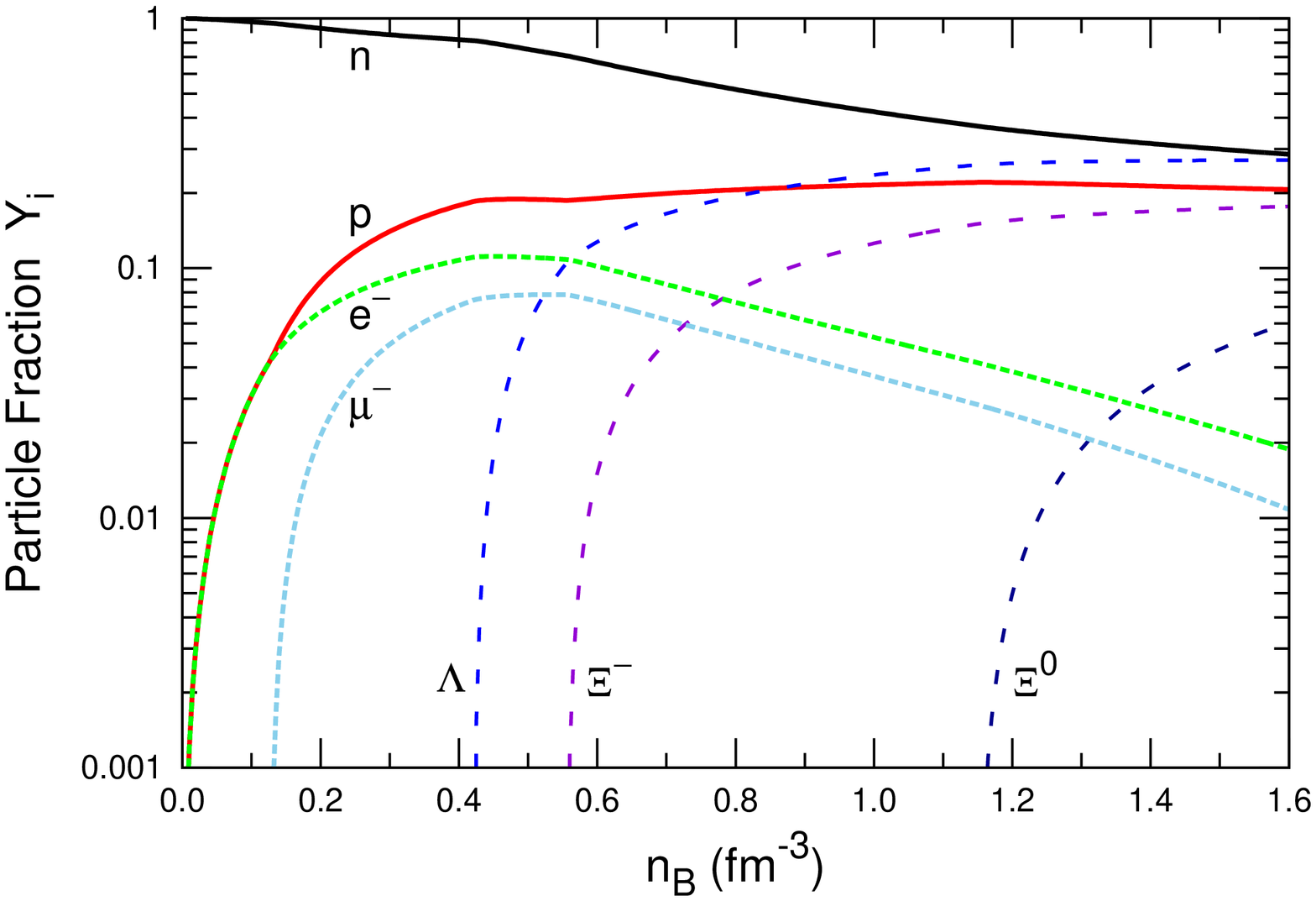}%
\caption{Particle fractions in a neutron star as functions of the total baryon density, $n_{B}$. 
The left (right) panel is for the case of SU(6) (beyond SU(6)) in the CQMC model.
}
\label{fig:Composition}       
\end{figure}

The particle fractions in a neutron star are shown in Fig.~\ref{fig:Composition}. 
In the case of ``beyond SU(6)", 
the creation densities of $\Lambda$ and $\Xi$ become higher than in the case of SU(6).  
In contrast, $\Sigma$ and $\Delta$-isobars do not appear in a neutron star. 
In the CQMC model, due to the effects of the in-medium baryon structure variation  
and the quark-quark hyperfine interaction, 
the potential $\Delta$-isobars feel in matter is shallower than in the case of Quantum Hadrodynamics. 
These effects suppress the appearance of $\Delta$-isobars. 
%
%
%

\section{Summary}
\label{sec:summary}

We have applied the chiral quark-meson coupling model 
to the EOS for a neutron star. 
Using the vector-meson couplings based on flavor SU(3) symmetry, we have found that 
the maximum mass of a neutron star can reach the recently observed mass, $1.97 \pm 0.04 M_{\odot}$, 
even in relativistic Hartree approximation. 
Furthermore, we have examined how  
$\Delta$-isobars contribute to the EOS,  
and shown that, because of the baryon structure variation in matter and 
the quark-quark hyperfine interaction, they do not appear in a neutron star. 
We finally note that the Fock term considerably contributes to the nuclear symmetry energy ($a_4$)~\cite{Katayama:2012ge}. 
\begin{acknowledgements}
The authors (especially, T. M.) thank Myung-Ki Cheoun for supporting this research. 
\end{acknowledgements}


\begin{thebibliography}{99}
%
%
\bibitem{Demorest:2010bx}
Demorest, P., Pennucci, T., Ransom, S., Roberts, M., Hessels, J.: 
Shapiro Delay Measurement of A Two Solar Mass Neutron Star. 
Nature 467, 1081 (2010)
\bibitem{Nagai:2008ai}
Nagai, S., Miyatsu, T., Saito, K., Tsushima, K.: 
Quark-meson coupling model with the cloudy bag. 
Phys. Lett. B 666, 239 (2008)
\bibitem{Saito:2005rv}
Saito, K., Tsushima, K., Thomas, A.W.: 
Nucleon and hadron structure changes in the nuclear medium and impact on observables. 
Prog. Part. Nucl. Phys. 58, 1 (2007)
\bibitem{Miyatsu:2011bc}
Miyatsu, T., Katayama, T., Saito, K.: 
Effects of Fock term, tensor coupling and baryon structure variation on a neutron star. 
Phys. Lett. B 709, 242 (2012)
\bibitem{Katayama:2012ge}
Katayama, T., Miyatsu T., Saito, K.: 
EoS for massive neutron stars. 
arXiv:1207.1554 [astro-ph.SR]
\bibitem{Rijken:2010zzb}
Rijken, T.A., Nagels, M.M., Yamamoto, Y.: 
Baryon-baryon interactions: Nijmegen extended-soft-core models. 
Prog. Theor. Phys. Suppl. 185, 14 (2010)
\end{thebibliography}


\end{document}